\documentstyle[preprint,aps,epsf]{revtex}

\begin{document}

\hoffset-1cm
\draft

\preprint{TPR-96-08 (revised);\  hep-th/9606055}

\title{Three-Point Functions at Finite Temperature}

\author{M.E. Carrington\cite{address1} and U. Heinz}

\address{
   Institut f\"ur Theoretische Physik, Universit\"at Regensburg,\\
   D-93040 Regensburg, Germany
}

\date{\today}

\maketitle

\begin{abstract}

We study 3-point functions at finite temperature in the closed time 
path formalism. We give a general decomposition of the eight component 
tensor in terms of seven vertex functions. We derive a spectral 
representation for these seven functions in terms of two independent 
real spectral functions. We derive  relationships between the seven 
functions and obtain a representation of the vertex tensor that 
greatly simplifies calculations in real time.  

\end{abstract} 


\narrowtext

\section{Introduction}  
\label{sec1}

In this paper, we discuss the analytical structure of the three-point 
function at finite temperature. We consider bosonic fields and work in 
the real time formalism. Throughout this paper we use the Keldysh or 
closed time path (CTP) formalism \cite{Keld,Landau,Rammer}. The 
integration contour has one branch running from negative infinity to 
positive infinity, and a second branch running backwards from positive 
infinity to negative infinity. We call the top branch of this contour 
$C_1$ and the bottom branch $C_2$. The arguments of the fields can 
take values on either branch of this contour. This choice of contour 
then leads to a doubling of degrees of freedom.  
 
Historically, the real time formalism has been much less popular than 
the imaginary time formalism because of the mathematical difficulties 
associated with the doubling of degrees of freedom. However, in some 
cases calculations in the real time formalism are much simpler. In 
particular, the real time formalism is extremely useful when 
working with ultra soft external energy scales. Examination of the 
pole structure of the internal propagators allows one to identify 
infinite sets of graphs that contribute to the same order in 
perturbation theory. For example, in a theory with an interaction of 
the form $g\phi^3$, the ladder graphs contribute terms to the self 
energy of order $g^{2L}T^{1+L}/k^{L-1}$ when the number 
of loops $L$ is larger than 1 \cite{Smilga,Carr}. When the pole 
structure of all propagators is explicit it is easy to extract the 
piece of each graph that dominates when $k\sim g^2T$. In order to 
obtain complete results, all such contributions must be resummed.
This requires an efficient and economic set of calculational 
tools for real-time perturbation theory to which we wish to
contribute with the present paper.

The analytic structure of the three-point vertex at finite temperature
has been studied previously in several publications, both in imaginary 
time \cite{evans,taylor} and real time 
\cite{TimE,RandyK,Aurenche,Eijck,EKW94}. The real time formulation has the 
advantage that it avoids the need for analytic continuation to real 
frequencies which, for higher order $n$-point functions, can easily 
become a source of confusion. Its disadvantage is the occurrence of 
many different thermal components of the $n$-point functions which 
results from the doubling of degrees of freedom. The resulting 
proliferation of terms in real-time calculations has contributed 
considerably to the lack of popularity of the real-time formalism.  
In the actual calculations one finds, however, that a large number of 
terms cancel in the end. The vertex representations found in the 
previous papers \cite{TimE,RandyK,Aurenche,Eijck,EKW94} do not implement 
these cancellation mechanisms at all, or only incompletely. We derive 
here a different (though mathematically equivalent) decomposition
and spectral representation of the real-time three-point vertex at
finite temperature. We demonstrate that it leads to impressive 
simplifications in perturbative calculations by making the physical
cancellations explicit on an algebraic level in the first step of the 
calculation, thereby dramatically reducing the number of terms which 
need to be evaluated.  

\section{Single-particle propagator}
\label{sec2}

To establish our notation and for later use we first consider the  
single-particle propagator. In real time, the propagator has 
$2^2=4$ components, since each of the two fields can take values on 
either branch of the contour. Thus, the two-point function can be 
written as a $2 \times 2 $ matrix of the form 
 \begin{equation}
 \label{2}
   D = \left(  \matrix {D_{11} & D_{12} \cr
                        D_{21} & D_{22} \cr} \right) \, ,
 \end{equation}
where $D_{11}$ is the propagator for fields moving along $C_1$, 
$D_{12}$ is the propagator for fields moving from $C_1$ to $C_2$, etc.  
The four components are given by \cite{LvW87}
 \begin{eqnarray}
 \label{eq: compD}
   D_{11}(x-y) &=& -i\langle T(\phi(x) \phi(y))\rangle \, , \nonumber\\ 
   D_{12}(x-y) &=& -i\langle \phi(y) \phi(x) \rangle \, , \nonumber\\
   D_{21}(x-y) &=& -i\langle \phi(x) \phi(y)\rangle \, , \nonumber\\
   D_{22}(x-y) &=& -i\langle\tilde{T}(\phi(x)\phi(y))\rangle \, , 
 \end{eqnarray}
where $T$ is the usual time ordering operator, and $\tilde{T}$ is the 
antichronological time ordering operator. These four components 
satisfy
 \begin{equation}
 \label{3}
   D_{11} - D_{12} - D_{21} + D_{22} = 0 
 \end{equation} 
as a consequence of the identity $\theta(x) + \theta(-x) =1$.  

It is more useful to write the propagator in terms of the three 
functions 
 \begin{eqnarray} \label{3a}
   D_R &=& D_{11} - D_{12} \, , \nonumber\\
   D_A &=& D_{11} - D_{21} \, , \nonumber\\ 
   D_F &=& D_{11} + D_{22} \, .
 \end{eqnarray}
$D_R$ and $D_A$ are the usual retarded and advanced propagators 
satisfying 
 \begin{equation}
 \label{4} 
   D_R(x-y)-D_A(x-y) = -i\langle [\phi(x),\phi(y)] \rangle\, ,
 \end{equation}
and $D_F$ is the symmetric combination
 \begin{equation}
 \label{4a} 
   D_F(x-y) = -i\langle \{\phi(x),\phi(y)\} \rangle\, .
 \end{equation}
In Appendix~\ref{appa1} we show that in momentum space these 
propagators are related by the well-known fluctuation-dissipation 
theorem \cite{FDT} 
 \begin{equation}
 \label{eq: 5}
   D_F(p) = \Bigl(1+2n(p_0)\Bigr)\, \Bigl(D_R(p) - D_A(p)\Bigr)
          = -i \, \Bigl(1+2n(p_0)\Bigr)\, \rho_-(p)\, . 
 \label{eq: KMSD}
 \end{equation}
where $n(p_0)$ is the thermal Bose-Einstein distribution 
 \begin{equation}
 \label{6}
   n(p_0) = \frac{1}{e^{\beta p_0}-1}. 
 \end{equation}
$\rho_-(p)$ is the spectral density in terms of which all components 
of the single-particle propagator can be expressed via spectral 
integrals, for example 
 \begin{equation}
 \label{eq: spectralprop}
   D_{R,A}(p) = \int_{-\infty}^\infty \frac{d\omega}{2\pi}
          {\rho_-(\omega,{\bf p}) \over
           p_0 - \omega \pm i\epsilon} \, .
 \end{equation}
Equations~(\ref{3}) and~(\ref{3a}) are inverted by
 \begin{eqnarray}
 \label{7}
   D_{11} &=& \frac{1}{2} (D_F + D_A + D_R) \, , \nonumber\\
   D_{12} &=& \frac{1}{2} (D_F  +D_A - D_R) \, , \nonumber\\
   D_{21} &=& \frac{1}{2} (D_F  -D_A + D_R) \, , \nonumber\\
   D_{22} &=& \frac{1}{2} (D_F  -D_A - D_R) \, . 
 \end{eqnarray}
These equations can be written in a more convenient notation as 
\cite{Chou}
 \begin{equation}
 \label{eq: decompD1}
   2\,D = D_R {1\choose 1}{1\choose -1} 
        + D_A {1\choose -1}{1\choose 1} 
        + D_F {1\choose 1}{1\choose 1}
 \end{equation} 
where the outer product of the column vectors is to be taken. Using 
the relation Eq.~(\ref{eq: KMSD}) we further obtain \cite{PeterH}
 \begin{equation}
   D(p) = D_R(p) {1\choose 1} {1+n(p_0)\choose n(p_0)} 
        - D_A(p) {n(p_0)\choose 1+n(p_0)} {1\choose 1}\, .
 \label{eq: decompD2}
 \end{equation} 
For later use we give the decomposition of the inverse propagator:
 \begin{eqnarray}
   D^{-1}(p) = \frac{1}{D_R(p)}{ 1 \choose -1}{ 1+n(p_0) \choose -n(p_0)} 
               + \frac{1}{D_A(p)}{ -n(p_0) \choose 1+n(p_0) }{1\choose -1}. 
 \label{eq: inverseprop}
 \end{eqnarray}
These representations are very useful for doing calculations in real 
time.  

Similar relations can be obtained for the polarization tensor, although 
they are perturbatively less useful. The polarization tensor is the 1PI 
two-point function and is obtained by amputating the external legs from 
the single-particle propagator. The Dyson equation gives
 \begin{equation}
 \label{8} 
   iD(p) = iD_0(p) + iD_0(p) \, \bigl(-i\Pi(p)\bigr)\, iD(p) \, .
 \end{equation}
The analogues of Eqs.~(\ref{3}) and~(\ref{3a}) are 
 \begin{eqnarray}
   \Pi_R &=& \Pi_{11} + \Pi_{12} \, , \nonumber\\ 
   \Pi_A &=& \Pi_{11} + \Pi_{21} \, , \nonumber\\ 
   \Pi_F &=& \Pi_{11} + \Pi_{22} \, ,
 \label{eq: physpi}
 \end{eqnarray} 
and
 \begin{equation}
   \Pi_{11} + \Pi_{12} + \Pi_{21} + \Pi_{22} = 0 \, .
 \label{eq: circpi}
 \end{equation}
The analogues of Eqs.~(\ref{eq: decompD1}),~(\ref{eq: KMSD}) and~(\ref{eq: decompD2}) are
 \begin{eqnarray}
 \label{Pidecomp1}
   2\, \Pi (p) &=& \Pi_R(p) {1\choose -1} {1\choose 1} 
             + \Pi_A(p) {1\choose 1} {1\choose -1}
             + \Pi_F(p) {1\choose -1} {1\choose -1}\, ,
 \\
 \label{eq: KMSpi} 
    \Pi_F(p) &=& \Bigl( 1+2n(p_0) \Bigr)\, 
                 \Bigl( \Pi_R(p) - \Pi_A(p) \Bigr) \, ,
 \\
 \label{Pidecomp2}
   \Pi (p) &=& \Pi_R(p) {1\choose -1} {1+n(p_0)\choose -n(p_0)} 
             + \Pi_A(p) {-n(p_0)\choose 1+n(p_0)} {1\choose -1}\, .
 \end{eqnarray}

\section{Three-Point Function}
\label{sec3}

In the real time formalism, the three-point function has $2^3 = 8$ 
components. We denote these by $\Gamma_{abc}$ with $\{a,b,c = 1,2\}$ 
where, for example,
 \begin{eqnarray}
   \Gamma_{111}(x,y,z) 
    &=& \langle T(\phi(x)\phi(y)\phi(z))\rangle \, , \nonumber\\
   \Gamma_{112}(x,y,z) 
    &=& \langle\phi(z)\,T(\phi(x)\phi(y))\rangle \, , 
 \label{eq: compVer}
 \end{eqnarray}
etc. The full set of vertex functions can be found in 
Refs.~\cite{Chou,LvW87}. Only seven of these eight components are 
independent because of the identity
 \begin{equation} 
   \sum_{a,b,c=1}^2 (-1)^{a+b+c-3} \Gamma_{abc} =0 
 \label{eq: circVer}
 \end{equation}
which follows in the same way as Eq.~(\ref{3}) from $\theta(x) + \theta(-x) 
= 1$. The seven combinations that we use are defined as \cite{Chou} 
 \begin{eqnarray}
   \Gamma_{R}(x,y,z) &=& \Gamma_{111} - \Gamma_{112} 
                - \Gamma_{211} + \Gamma_{212} \, ,
 \nonumber\\ 
   \Gamma_{Ri}(x,y,z) &=& \Gamma_{111} - \Gamma_{112} 
                   - \Gamma_{121} + \Gamma_{122} \, ,
 \nonumber\\ 
   \Gamma_{Ro}(x,y,z) &=& \Gamma_{111} - \Gamma_{121} 
                   - \Gamma_{211} + \Gamma_{221} \, ,
 \nonumber\\ 
   \Gamma_{F}(x,y,z) &=& \Gamma_{111} - \Gamma_{121} 
                + \Gamma_{212} - \Gamma_{222} \, ,
 \nonumber\\
   \Gamma_{Fi}(x,y,z) &=& \Gamma_{111} + \Gamma_{122} 
                   - \Gamma_{211} - \Gamma_{222} \, ,
 \nonumber\\
   \Gamma_{Fo}(x,y,z) &=& \Gamma_{111} - \Gamma_{112} 
                   + \Gamma_{221} - \Gamma_{222} \, ,
 \nonumber\\
   \Gamma_{E}(x,y,z) &=& \Gamma_{111} + \Gamma_{122} 
                + \Gamma_{212} + \Gamma_{221} \, .
 \label{eq: physVer}
 \end{eqnarray} 
Clearly, the set 
Eq.~(\ref{eq: physVer}) contains the same information as 
Eqs.~(\ref{eq: compVer}) and~(\ref{eq: circVer}). It merely defines a 
change of basis to a different set of independent functions, just as 
Eq.~(\ref{3a}) defines a change of independent functions for the 
single-particle propagator.  

In coordinate space we always label the first leg of the three-point 
function by $x$ and call it the ``incoming leg $(i)$", the third leg 
we label by $z$ and call it the ``outgoing leg $(o)$", and the second 
(middle) leg we label by $y$. Using Eq.~(\ref{eq: compVer}) we can 
rewrite Eq.~(\ref{eq: physVer}) to obtain expressions for the vertices in 
terms of commutators of the fields. With the obvious shorthands 
$\phi_1 \equiv \phi(x)$, $\phi_2 \equiv \phi(y)$, $\phi_3 \equiv 
\phi(z)$, and $\theta_{12} \equiv \theta(x_0-y_0)$, etc., one thus 
finds that the first three vertices in Eq.~(\ref{eq: physVer}) are just 
the retarded vertex functions:  
 \begin{eqnarray}
   \Gamma_{R} &=& 
     \theta_{23} \theta_{31} \langle [[\phi_2,\phi_3], \phi_1] \rangle 
   + \theta_{21} \theta_{13} \langle [[\phi_2,\phi_1], \phi_3] \rangle \, ,
 \nonumber\\ 
   \Gamma_{Ri} &=& 
     \theta_{12} \theta_{23} \langle [[\phi_1,\phi_2], \phi_3] \rangle 
   + \theta_{13} \theta_{32} \langle [[\phi_1,\phi_3], \phi_2] \rangle \, ,
 \nonumber\\ 
   \Gamma_{Ro} &=& 
     \theta_{32} \theta_{21} \langle [[\phi_3,\phi_2], \phi_1] \rangle 
   + \theta_{31} \theta_{12} \langle [[\phi_3,\phi_1], \phi_2] \rangle \, .
 \label{commutators}
 \end{eqnarray} 
$\Gamma_{Ri}$ is the vertex retarded with respect to $x_0$, 
$\Gamma_{Ro}$ is retarded with respect to $z_0$, and $\Gamma_R$ is 
retarded with respect to $y_0$. The four remaining vertices contain 
various symmetric combinations of the fields.  

The inversion of Eq.~(\ref{eq: physVer}) can be elegantly presented 
in tensor notation in terms of the two-component column vectors
introduced \cite{Chou} in the previous Section:   
 \begin{eqnarray}
   4\,\Gamma &=& \Gamma_R {1 \choose -1} {1\choose 1} {1\choose -1}
               + \Gamma_{Ri} {1 \choose 1}{1\choose -1} {1 \choose -1}
 \nonumber\\
           &+& \Gamma_{Ro} {1 \choose -1} {1\choose -1} {1 \choose 1}
             + \Gamma_F {1\choose 1} {1\choose -1}{1\choose 1}
 \nonumber\\
           &+& \Gamma_{Fi} {1\choose -1}{1\choose 1}{1\choose 1}
             + \Gamma_{Fo} {1\choose 1}{1\choose 1}{1\choose -1}
 \nonumber\\
           &+& \Gamma_E {1\choose 1}{1\choose 1}{1\choose 1}.
 \label{eq: decompver}
 \end{eqnarray}
This decomposition of the thermal vertex tensor is analogous to 
Eq.~(\ref{eq: decompD1}) for the propagator. Its simple structure 
results from our choice Eq.~(\ref{eq: physVer}) for the basis functions.  
In the following Section we show that it is possible to further 
simplify this decomposition by deriving relationships between the 
vertex functions (in analogy to Eq.~(\ref{eq: 5})). The usefulness of 
these expressions for performing explicit calculations is demonstrated 
in Sec.~\ref{sec4}.  

\subsection{Spectral representation of the three-point functions}
\label{sec31}

In Appendix~\ref{appa2} we show in detail that the seven three-point 
functions Eq.~(\ref{eq: physVer}) can be rewritten in terms of two 
independent spectral densities. Starting from (\ref{commutators}), the 
three-point functions are constructed as products of theta functions 
and expectation values of the form 
 \begin{eqnarray}
   \langle \phi_a\phi_b\phi_c \rangle 
   = \sum_l \langle l | e^{-\beta H} \phi_a\phi_b\phi_c | l \rangle \, .
 \label{24}
 \end{eqnarray}
We calculate these expectation values in the usual way, by going to 
the Heisenberg representation and using translation invariance in the 
form of Eq.~(\ref{A1}), and inserting complete sets of states between the 
operators.  Because of the  cyclic property of the trace, there are 
only two independent expectation values. This allows us to write all 
of the three-point functions in terms of two spectral functions. We 
find it useful to define 
 \begin{eqnarray}
  \rho_A = \rho_1+\rho_3
 \nonumber\\
  \rho_B = \rho_2+\rho_4
 \label{AB}
 \end{eqnarray}
where
 \begin{eqnarray}
  \rho_1(p,k) &=& (2\pi)^8 \sum_{lmn} M_{nml} 
  (e^{-\beta E_l} - e^{-\beta E_m}) \, 
  \delta(p - {\textstyle{1\over 2}}k + p_l - p_n)\, 
  \delta(p + {\textstyle{1\over 2}}k + p_l - p_m) \, ,
 \nonumber\\
  \rho_2(p,k) &=& (2\pi)^8  \sum_{lmn} M_{nml}
  (e^{-\beta E_n} - e^{-\beta E_m}) \,
  \delta(p - {\textstyle{1\over 2}}k + p_l - p_n)\, 
  \delta(p + {\textstyle{1\over 2}}k + p_l - p_m) \, ,
 \nonumber\\
  \rho_3(p,k) &=&  (2\pi)^8 \sum_{lmn} M_{nml}
  (e^{-\beta E_l} - e^{-\beta E_n}) \,
  \delta(p - {\textstyle{1\over 2}}k + p_m - p_l)\, 
  \delta(p + {\textstyle{1\over 2}}k + p_n - p_l) \, ,
 \nonumber\\
  \rho_4(p,k) &=&  (2\pi)^8 \sum_{lmn} M_{nml}
  (e^{-\beta E_m} - e^{-\beta E_n}) \,
  \delta(p - {\textstyle{1\over 2}}k + p_m - p_l) \,
  \delta(p + {\textstyle{1\over 2}}k + p_n - p_l) \, ,
 \label{eq: spectralfunctions} 
 \end{eqnarray}
with $M_{nml} = \langle l|\phi(0) |n\rangle \langle n|\phi(0)|m\rangle 
\langle m|\phi(0) |l\rangle = M^*_{lmn}$. The two independent spectral 
functions $\rho_A$ and $\rho_B$ are real in coordinate space since 
 \begin{eqnarray}
 \label{real}
   \rho_{A,B}(p,k) = \rho^*_{A,B}(-p,-k).
 \end{eqnarray}
Another way of verifying that there are only two independent 
real spectral densities is as follows: all four spectral functions
$\rho_i(p,k)$, $i=1,\dots,4$, in Eq.~(\ref{eq: spectralfunctions}) can be
expressed in terms of a single complex function,
 \begin{eqnarray}
  \rho(p,k) &=& (2\pi)^8 \sum_{lmn} M_{nml} 
  e^{-\beta E_l} \, 
  \delta(p - {\textstyle{1\over 2}}k + p_l - p_n)\, 
  \delta(p + {\textstyle{1\over 2}}k + p_l - p_m) \, ,
 \label{spectral1}
 \end{eqnarray}
by expressing all energies in the Boltzmann factors in terms of $E_l$, 
using the $\delta$-functions, and appropriately relabelling the 
summation indices.  So there is really only one complex function (or, 
equivalently, two real functions) of $p$ and $k$ which contain all the 
physical information on the (non-perturbative) analytical structure of 
the vertex.  

We now write down the spectral representation of the vertex functions
in momentum space, $\Gamma_\alpha(p_1,p_2,p_3)$, where $p_1, p_2, p_3$ 
(with $p_i = (E_i, {\bf p}_i)$) are the momenta associated with 
$x,y,z$, respectively (i.e. the three momenta flowing into the three 
legs of the vertex). Due to translation invariance $p_1+p_2+p_3=0$, and
we exploit this by writing $p_1=p-k/2$, $p_2 = k$, and $p_3 = -(p+k/2)$. 

We use the following shorthand notation:
 \begin{eqnarray}
 \label{short}
    \Gamma_\alpha &\equiv& \Gamma_\alpha(p,k) \equiv 
                           \Gamma_\alpha(p-k/2,k,-p-k/2)
    = \Gamma_\alpha(p_1,p_2,p_3) \, , \quad \alpha=R,F,E,\dots\, ;
 \nonumber\\
    \rho_{A,B} &\equiv& \rho_{A,B}(\omega_1,{\bf p};\omega_2,{\bf k})
    \, ;
 \nonumber\\
    a_j^\pm &=& {1 \over E_j - \Omega_j \pm i\epsilon}\, , \quad j=1,2,3\, ;
 \nonumber\\
    \Omega_1 &=& \omega_1 - {\textstyle{1\over 2}}\omega_2\, , \quad
    \Omega_2 = \omega_2\, ,\quad
    \Omega_3 = - \omega_1 - {\textstyle{1\over 2}}\omega_2\, , \quad
    \Omega_1 + \Omega_2 + \Omega_3 =0 \, ;
 \nonumber\\
    \tilde N_i &=& 1 + 2 n(\Omega_i)\, , \quad i=1,2,3 \, ;
 \nonumber\\
    \int &=& \int_{-\infty}^{\infty}
             \frac{d\omega_1}{2\pi}\frac{d\omega_2}{2\pi} \, .
 \end{eqnarray}
Then the spectral integrals for the vertex functions are
 \begin{eqnarray}
 \label{eq: spectral} 
   \Gamma_R     &=& \int \left[ (\rho_B-\rho_A) a_1^- - \rho_A a_3^- \right]
                    a_2^+ \, ,
 \nonumber\\
   \Gamma_{Ri} &=& \int \left[ \rho_A a_3^- + \rho_B a_2^- \right]
                    a_1^+ \, ,
 \nonumber\\
   \Gamma_{Ro} &=& \int \left[ (\rho_A-\rho_B) a_1^- - \rho_B a_2^- \right]
                    a_3^+ \, ,
 \nonumber\\
   \Gamma_F     &=& \int \left[ \tilde N_1 (\rho_B-\rho_A) 
                                (a_1^+ a_2^- + a_1^- a_3^+ + a_2^- a_3^+)
                              - \tilde N_3 \rho_A
                                (a_1^+ a_2^- + a_1^+ a_3^- + a_2^- a_3^+)
                         \right] \, ,
 \nonumber\\
   \Gamma_{Fi} &=& \int \left[ \tilde N_3 \rho_A 
                                (a_1^- a_2^+ + a_1^- a_3^+ + a_2^+ a_3^-)
                              + \tilde N_2 \rho_B
                                (a_1^- a_2^+ + a_1^- a_3^+ + a_2^- a_3^+)
                         \right] \, ,
 \nonumber\\
   \Gamma_{Fo} &=& \int \left[ \tilde N_1 (\rho_A-\rho_B) 
                                (a_1^- a_2^+ + a_1^+ a_3^- + a_2^+ a_3^-)
                              - \tilde N_2 \rho_B
                                (a_1^+ a_2^- + a_1^+ a_3^- + a_2^+ a_3^-)
                         \right] \, ,
 \nonumber\\
   \Gamma_E     &=& \int \Bigl[ \Bigl( \tilde N_1 \tilde N_3 (\rho_B-\rho_A)
                                     + \tilde N_2 \tilde N_3 \rho_B \Bigr)
                                (a_1^+ + a_2^+) a_3^-
                - \tilde N_1 (\tilde N_3 \rho_A + \tilde N_2 \rho_B)
                                (a_2^+ + a_3^+) a_1^-
 \nonumber\\
           && \quad \quad + \Bigl( \tilde N_1 \tilde N_2 (\rho_A-\rho_B)
                                 + \tilde N_2 \tilde N_3 \rho_A \Bigr)
                                (a_1^+ + a_3^+) a_2^-  \Bigr] \, .
 \end{eqnarray}
These are the real-time analogues of the spectral representations
for the vertex function given by Evans in the imaginary time formalism
(ITF) \cite{evans}. Explicit expressions for the spectral densities in 
the ``hard thermal loop approximation'' were given in Ref.~\cite{taylor} 
for the ITF vertex; their transcription into the real time formalism
will be an interesting subject for a separate publication. 

\subsection{Relationships between the three-point functions}
\label{sec32}
 
From the spectral representation, we obtain relationships between 
the three-point functions which allow us to eliminate the three 
$\Gamma_F$'s and $\Gamma_E$ in terms of the retarded vertex functions
and their complex conjugates. We use again shorthands $N_i = N(E_i) = 
1 + 2 n(E_i) = 1 + 2 n_i$ where $E_1, E_2, E_3$ are the 
energies flowing into the three legs of the vertex.  Using
 \begin{eqnarray}
    {\rm Re} \,a_j^{\pm} &=& \frac{1}{E_j-\Omega_j},
 \nonumber\\
    {\rm Im} \,a_j^{\pm} &=& \mp i\pi\delta(E_j-\Omega_j),
 \label{ReIm}
 \end{eqnarray}
and doing some tedious algebra we obtain  
 \begin{eqnarray}
   \Gamma_F &=& N_1 (\Gamma^*_R - \Gamma_{Ro})
              + N_3 (\Gamma^*_R - \Gamma_{Ri}) \, ,
 \nonumber\\
   \Gamma_{Fi} &=& N_2 (\Gamma^*_{Ri} - \Gamma_{Ro}) 
                  + N_3 (\Gamma^*_{Ri} - \Gamma_R) \, ,
 \nonumber\\
   \Gamma_{Fo} &=& N_1 (\Gamma^*_{Ro} - \Gamma_R) 
                  + N_2 (\Gamma^*_{Ro} - \Gamma_{Ri}) \, ,
 \nonumber\\
   \Gamma_E &=& \Gamma^*_{Ri} + \Gamma^*_R + \Gamma^*_{Ro}
              + N_2 N_3 (\Gamma_{Ri} + \Gamma^*_{Ri})
 \nonumber\\
            &+& N_1 N_3 (\Gamma_R + \Gamma^*_R)
              + N_1 N_2 (\Gamma_{Ro} + \Gamma^*_{Ro}) \, , 
 \label{eq: KMSver}
 \end{eqnarray}
where we have used the identity
 \begin{equation}
   N_1 N_2 + N_2 N_3 + N_3 N_1 = -1 \, . 
 \end{equation}

Substituting Eq.~(\ref{eq: KMSver}) into Eq.~(\ref{eq: decompver}) we obtain the 
analogue of Eq.~(\ref{eq: decompD2}):
 \begin{eqnarray}
   \Gamma &=&  \Gamma_R {n_1\choose 1+n_1} {1\choose 1} {n_3\choose 1+n_3} 
             -\frac{1}{2}\Gamma^*_R(N_1+N_3)
              {1\choose 1}{n_2\choose 1+n_2}{1\choose 1}
 \nonumber\\
          &+& \Gamma_{Ri} {1\choose 1} {n_2 \choose 1+n_2} {n_3\choose 1+n_3} 
            - \frac{1}{2} \Gamma_{Ri}^* (N_2+N_3)
              {n_1\choose 1+n_1}{1\choose 1}{1\choose 1}
 \nonumber\\
          &+& \Gamma_{Ro} {n_1\choose 1+n_1} {n_2\choose 1+n_2} {1\choose 1} 
             -\frac{1}{2}\Gamma_{Ro}^* (N_1+N_2)
              {1\choose 1}{1\choose 1}{n_3\choose 1+n_3}.
 \label{eq: DECOMP}
 \end{eqnarray}

This decomposition, together with the spectral representations given 
by the first three lines in Eq. (\ref{eq: spectral}) which express the 
three (complex) retarded vertices in terms of two (real) spectral 
densities, is the main result of this paper. It is mathematically 
equivalent to, but structurally simpler than the results obtained by 
Evans \cite{TimE}, Kobes \cite{RandyK}, Aurenche and Becherrawy 
\cite{Aurenche}, and van Eijck, Kobes, and van Weert \cite{Eijck,EKW94}. 
The focus of those papers was the comparison of the imaginary-time and 
real-time formalisms, and transformations between various representations 
of the real-time formulation. Relations were obtained between the 
individual components of the vertex tensor and the retarded and 
advanced vertices. The basic procedure in all cases was similar: 
One started with the observation that the propagator can be 
diagonalized by a simple matrix transformation. This diagonalization 
procedure is not unique, resulting in different formulations using 
different basis functions (retarded/advanced \cite{Aurenche}, $F/\bar F$
\cite{EKW94}, etc., see \cite{Gelis} for a recent review). One then 
attempts to write the vertex tensor in a similar way, as a simple 
core function contracted with a similar matrix transformation at each 
leg \cite{TimE}. Unfortunately, none of the previously obtained 
results is particularly simple.  

The key to our derivation is the realization that the transformations 
between the various (anti)symmetric vertices Eq.~(\ref{eq: physVer}) are 
much simpler than transformations involving the individual components 
of the vertex tensor $\Gamma_{abc}$. In this sense, the 
(anti)symmetric combinations that we use are more physical than, e.g., 
the time-ordered vertex $\Gamma_{111}$. Our result Eq.~(\ref{eq: DECOMP}), 
expressed in the column vector notation of \cite{PeterH}, is extremely 
simple to use in calculations, as will be demonstrated in the next 
section.    

\section{Outline of a calculation with full vertices} 
\label{sec4}

As an example of the usefulness of these techniques, in particular the 
decomposition Eq.~(\ref{eq: DECOMP}), we set up the calculation of the 
polarization tensor in $\phi^3$ theory in six dimensions. This 
renormalizable theory is interesting as a toy model since the $\phi^3$ 
interaction is mathematically similar to the three gluon interaction 
in QCD. The Lagrangian is given by 
 \begin{eqnarray}
   {\cal L} =\frac{1}{2} (\partial_\mu \phi)^2 - \frac{1}{2}m^2\phi^2 
            -\frac{g}{3!}\phi^3.
 \label{Lag}
 \end{eqnarray}
In terms of full propagators and vertices the polarization tensor is 
given by the 1-loop diagram shown in 
Fig. 1.

In order for the polarization tensor and vertex to satisfy the coupled 
Schwinger-Dyson equations, one must use one corrected vertex and 
one bare vertex. We follow the notation of \cite{PeterH}. The bare 
vertex contributes a factor of $-ig\tau^3$, where $\tau^3$ is the third 
Pauli matrix. We obtain
 \begin{eqnarray}
    \Pi_{ab}(k) = g\int \frac{d^6p}{(2\pi)^6}
                  \tilde{\Gamma}_{cad}(p-k/2,k,-(p+k/2)) \,
                  D_{bd}(-(p+k/2))\, \tau^3_{bx} \, D_{xc}(p-k/2)
 \end{eqnarray}
where $\tilde{\Gamma}$ indicates the full 1PI vertex. We rewrite this 
equation in terms of the connected vertex $\Gamma$ by using the definition 
 \begin{equation}
   \tilde{\Gamma}_{cad}(p_1,p_2,p_3) = \frac{1}{i^3}\,
   D^{-1}_{cc'}(p_1) \, D^{-1}_{aa'}(p_2) \, D^{-1}_{dd'}(p_3) \,
   \Gamma_{c'a'd'}(p_1,p_2,p_3)
 \end{equation}
which gives 
 \begin{equation}
\label{eq: xxx}
   \Pi_{ab}(k) = i g\, D^{-1}_{aa'}(k)\, \tau^3_{bx} 
   \left(\int \frac{d^6p}{(2\pi)^6}\Gamma_{xa'b}(p-k/2,k,-(p+k/2)) \right)\, .
\end{equation}

We insert the decompositions Eqs.~(\ref{eq: inverseprop}) and~(\ref{eq: 
DECOMP}) in terms of outer products of 2-component column vectors and 
use the rule \cite{PeterH} that for internal indices 
($a^{\prime},c^{\prime},d^{\prime},c,d,x$) which are to be summed over 
the corresponding two column vectors in Eqs. (\ref{eq: inverseprop}) 
and~(\ref{eq: DECOMP}) must be contracted to a scalar:
 \begin{equation}
 \label{rule}
   {x_1 \choose x_2} \cdot {x_3 \choose x_4} = x_1 x_3 + x_2 x_4 \, .
 \end{equation}
The index $b$ in Eq.~(\ref{eq: xxx}) is not summed over; the rule \cite{PeterH} is that, after multiplying the $\tau_3$-matrix into the vertex by summation over the index $x$, the remaining two column vectors with the index $b$ are to be contracted to a new column vector according to
\begin{equation}
 {x_1 \choose x_2} {x_3 \choose x_4} ={x_1   x_3 \choose x_2   x_4} \, .
\end{equation}

Since Eq.~(\ref{eq: DECOMP}) contains 6 terms and Eq.~(\ref{eq: 
inverseprop}) contains 2 terms, one expects 12 separate terms in the 
polarization tensor each of which is again a sum of two terms 
according to Eq.~(\ref{rule}). It is, however, known from practical 
experience that many of these terms cancel. In our formulation
this cancellation occurs at the very first step of the calculation
because many of the contractions Eq.~(\ref{rule}) vanish identically, e.g.
 \begin{equation} 
  {1 \choose 1} \cdot {1\choose -1} = 0\, , \qquad
  {1+n \choose -n} \cdot {n\choose 1+n} = 0\, .
 \end{equation}
Furthermore, when calculating, for example, the retarded polarization 
operator, all terms vanish in which the column vector carrying the 
index $b$ is given by ${1\choose -1}$.  

One thus obtains directly the following simple expressions:
 \begin{eqnarray} 
 \label{piresult}
   \Pi_R(k) &=& \Pi_{11}(k) + \Pi_{12}(k)
 \nonumber\\
    &=& ig\frac{1}{D_R(k)}\int \frac{d^6p}{(2\pi)^6}
        N_1 (\Gamma_{Ro}^* - \Gamma_R)  \, ,
 \nonumber\\
   \Pi_A(k) &=& \Pi_{11}(k) + \Pi_{21}(k)
 \nonumber\\
   &=&  ig\frac{1}{D_A(k)}\int \frac{d^6p}{(2\pi)^6}
       N_1(\Gamma_R^*-\Gamma_{Ro}) = \Pi_R^*(k) \, ,
 \nonumber\\
    \Pi_F(k)&=& \Pi_{11}(k)+\Pi_{22}(k)
 \nonumber\\
    &=& ig\int \frac{d^6p}{(2\pi)^6} N_1 N_2 
        \left[\frac{1}{D_R(k)}(\Gamma_{Ro}^*-\Gamma_{R}) 
            - \frac{1}{D_A(k)}(\Gamma_R^*-\Gamma_{Ro})\right] \, .
 \end{eqnarray}
The only additional input were the relations Eqs.~(\ref{eq: circpi}) and~(\ref{eq: KMSpi}) which result in the simple identities
 \begin{eqnarray}
 \label{piident}
   \int \frac{d^6p}{(2\pi)^6} (\Gamma_{Ri} + \Gamma_{Ro}) &=& 0, 
 \nonumber \\
   \int \frac{d^6p}{(2\pi)^6} (\Gamma_R + \Gamma_R^*) &=& 0. 
 \end{eqnarray}

The above results express the components of the full scalar self 
energy in terms of the full propagators and vertices and are thus 
correct to all orders of perturbation theory. Formally, the 
expressions in Eq.~(\ref{piresult}) and the identities 
Eq. (\ref{piident}) remain unchanged in $D$ space-time dimensions (with 
arbitrary $D$). The simple structure given in Eq.~(\ref{piresult}) 
should be preserved in partially resummed approximation schemes like 
e.g. the hard thermal loop approximation, in order to remain 
consistent with the Schwinger-Dyson equations.  

\section{Conclusions}
\label{sec5}

We have given a simple decomposition of the eight component real time 
vertex tensor in terms of seven vertex functions and outer products of 
triplets of simple two-component column vectors. We have shown that 
the seven vertex functions can be obtained through a spectral 
representation from two independent real spectral functions. We have 
obtained a set of relationships between the seven vertex functions 
which allows us to express all of them through the three retarded 
vertex functions and their complex conjugates. These expressions were 
used to derive another vector decomposition of the vertex tensor in 
terms of the retarded vertex functions and their complex conjugates.  
The advantage of this expression for performing calculations is that 
it allows one to see immediately that, after performing all of the 
contractions, many of the terms give zero. As far as we can see, our 
formalism implements all of the known cancellations between 
contributions from different components of the propagator and vertex 
tensors algebraically through the contraction rules for the column 
vectors. These cancellations occur before any loop integrals must be
evaluated, at the very beginning of the calculation. This greatly 
simplifies loop calculations in real-time finite temperature field 
theory.  

We note that work similar to that presented here was recently done by 
Fauser and Henning for the Yukawa theory \cite{PeterH2}.

\acknowledgments

M.E.C. gratefully acknowledges support by the Alexander von Humboldt 
Foundation through a Research Fellowship. U.H. would like to thank 
B.~M\"uller and the Physics Department at Duke University, were a large 
fraction of his work was done during a sabbatical leave, for their 
warm hospitality and stimulating discussions. The work of U.H. was 
supported by DFG, BMBF, and GSI.  

\appendix
\section{Lehmann spectral representations}
\label{appa}
\subsection{Single-particle propagators}
\label{appa1}

We give a short derivation of Eq.~(\ref{eq: KMSD}). We start from 
Eqs.~(\ref{4}) and (\ref{4a}) and evaluate the thermal expectation 
value in the Heisenberg picture by inserting complete sets of energy 
and momentum eigenstates. Exploiting translation invariance in the 
form 
 \begin{equation}
 \label{A1}
   \phi(x) = e^{ip\cdot x} \phi(0) e^{-ip\cdot x} \, ,
 \end{equation}
where $p$ is the momentum (gradient) operator and 
$\phi(0)$ the Schr\"odinger field operator, we get
 \begin{eqnarray}
 \label{A2}
   \langle [\phi(x)\phi(y)]_\pm \rangle &=&
     \text{Tr} \left( e^{-\beta H} [\phi(x)\phi(y)]_\pm 
               \right)
 \nonumber\\
   &=& \sum_{m,n} e^{-\beta E_m} 
       \left\vert \langle m \vert \phi(0) \vert n \rangle \right\vert^2
       \left( e^{i(p_m-p_n)\cdot(x-y)} \pm e^{i(p_m-p_n)\cdot(y-x)}
       \right) \, ,
 \end{eqnarray}
where $p^0_{m,n}=E_{m,n}$. Fourier transforming with respect to $x-y$ 
one finds
 \begin{eqnarray}
 \label{A3}
   \rho_\pm(p) &\equiv & \int d^4(x-y)e^{ip(x-y)}
   \langle [\phi(x),\phi(y)]_{\pm}\rangle
 \nonumber\\
   &=& (2\pi)^4 \sum_{m,n} \vert \phi_{mn}(0) \vert^2\, 
   \delta(p+p_m-p_n) \left( e^{-\beta E_m} \pm e^{-\beta E_n} \right)
   \, .
 \end{eqnarray}
Using the energy conserving $\delta$-function we can write
 \begin{equation}
 \label{A4}
   e^{-\beta E_m} + e^{-\beta E_n} = (1 + 2 n(p_0)) 
   \left( e^{-\beta E_m} - e^{-\beta E_n} \right)\, .
 \end{equation}
From Eqs.~(\ref{A3}), (\ref{A4}), (\ref{4}) and (\ref{4a}) we obtain 
 \begin{eqnarray}
   D_F(p) &=& -i\rho_+(p) = -i(1+2n(p_0))\rho_-(p)
 \nonumber\\
   &=& (1+2n(p_0))(D_R(p) - D_A(p)) 
 \nonumber
 \end{eqnarray}
in agreement with Eq.~(\ref{eq: KMSD}).

Alternatively, one can start from the KMS condition \cite{KMS} 
 \begin{equation}
 \label{KMS}
   \langle A({\bf x},x_0) B({\bf y},y_0) \rangle =
   \langle B({\bf y},y_0-i\beta) A({\bf x},x_0) \rangle =
   \langle B({\bf y},y_0) A({\bf x},x_0+i\beta) \rangle 
 \end{equation}
and apply it to $\langle \phi(x) \phi(y) \pm \phi(y) \phi(x) \rangle$. 
In momentum space this leads directly to
 \begin{equation}
 \label{A4a}
   \rho_+(p) = (1+2n(p_0)) \rho_-(p)
 \end{equation}
where $\rho_\pm$ are the respective Fourier transforms, with Lehmann 
representations Eq.~(\ref{A3}).

\subsection{Three point functions}
\label{appa2}

In this section we show in more detail how to obtain the spectral 
representation for the seven connected vertex functions.  We will 
show that the seven vertex functions can be written in terms of 
two independent spectral functions. We define 
 \begin{eqnarray} 
        a_1 &=& \phi_1\phi_2\phi_3 \theta_{12}\theta_{23},
 \nonumber\\
        b_1 &=& \phi_3\phi_1\phi_2 \theta_{12}\theta_{23},
 \nonumber\\
        c_1 &=& \phi_3\phi_2\phi_1 \theta_{12}\theta_{23},
 \nonumber\\
        d_1 &=& \phi_2\phi_1\phi_3 \theta_{12}\theta_{23}.
 \label{A5} 
 \end{eqnarray}
We have similar definitions for other combinations of theta functions. 
Using an obvious short hand notation we have  
 \begin{eqnarray} 
   \{a,b,c,d\}_2 &=& 
   \{\phi_1\phi_2\phi_3, \phi_2\phi_3\phi_1, 
     \phi_3\phi_2\phi_1, \phi_1\phi_3\phi_2\}\theta_{32}\theta_{21},
 \nonumber\\
   \{a,b,c,d\}_3 &=& 
   \{\phi_1\phi_2\phi_3, \phi_2\phi_3\phi_1, 
     \phi_3\phi_2\phi_1, \phi_1\phi_3\phi_2\}\theta_{23}\theta_{31},
 \nonumber\\
   \{a,b,c,d\}_4 &=&
   \{\phi_1\phi_2\phi_3, \phi_3\phi_1\phi_2, 
     \phi_3\phi_2\phi_1, \phi_2\phi_1\phi_3\}\theta_{21}\theta_{13},
 \nonumber\\
   \{a,b,c,d\}_5 &=&
   \{\phi_2\phi_3\phi_1, \phi_3\phi_1\phi_2, 
     \phi_1\phi_3\phi_2, \phi_2\phi_1\phi_3\}\theta_{13}\theta_{32},
 \nonumber\\
   \{a,b,c,d\}_6 &=& 
   \{\phi_2\phi_3\phi_1, \phi_3\phi_1\phi_2, 
     \phi_1\phi_3\phi_2, \phi_2\phi_1\phi_3\}\theta_{31}\theta_{12}. 
 \label{A6}
 \end{eqnarray}

In terms of these definitions, the seven vertex functions are 
 \begin{eqnarray} 
   \Gamma_R &=& -(a_3 - b_3 + c_3 - d_3) - (a_4 - b_4 + c_4 - d_4),
 \nonumber\\
   \Gamma_{Ri} &=& (a_1 - b_1 + c_1 - d_1) + (a_5 - b_5 + c_5 - d_5),
 \nonumber\\
   \Gamma_{Ro} &=& (a_2 - b_2 + c_2 - d_2) - (a_6 - b_6 + c_6 - d_6),
 \nonumber\\
   \Gamma_F &=& (a_1 + b_1 - c_1 - d_1) - (a_2 + b_2 - c_2 - d_2) 
              - (a_5 - b_5 - c_5 + d_5) - (a_6 - b_6 - c_6 + d_6),
 \nonumber\\
   \Gamma_{Fi} &=&(b_2 - a_2 + c_2 - d_2) - (a_3 - b_3 - c_3 + d_3) 
                   - (a_4 + b_4 - c_4 - d_4) + (a_6 + b_6 - c_6 - d_6),
 \nonumber\\
   \Gamma_{Fo} &=& (a_1 - b_1 - c_1 + d_1) + (a_3 + b_3 - c_3 - d_3) 
                 + (a_4 - b_4 - c_4 + d_4) - (a_5 + b_5 - c_5 - d_5),
 \nonumber\\
   \Gamma_E &=& (a_1+b_1+c_1+d_1) + (a_2+b_2+c_2+d_2) + (a_3+b_3+c_3+d_3) 
              + (a_4+b_4+c_4+d_4)
 \nonumber\\
   &\,&\,\, + (a_5+b_5+c_5+d_5) + (a_6+b_6+c_6+d_6). 
 \label{A7}
 \end{eqnarray}
We calculate the expectation values of the form Eq.~(\ref{24}) by using
Eq.~(\ref{A1}), inserting complete sets of states between the operators 
and using the cyclic property of the trace.  We obtain the four 
spectral functions Eq.~(\ref{eq: spectralfunctions}). 
 
Using the notation of Eq.~(\ref{short}), the 24 functions Eqs.~(\ref{A5}) and~(\ref{A6}) can be written in terms of the four spectral functions 
(\ref{eq: spectralfunctions}) as  
 \begin{eqnarray}
   \left\{ \matrix{ (a_1-b_1)\cr (a_1 + b_1)\cr (c_1-d_1)\cr 
                   -(c_1+d_1)\cr } \right\} &=&  
   \int_{-\infty}^{\infty}\frac{d\omega_1}{2\pi}\frac{d\omega_2}{2\pi} 
   a_1^+a_3^- 
   \left\{ \matrix{ \rho_1 \cr -\tilde{N}_3 \rho_1 \cr \rho_3 \cr 
                   -\tilde{N}_3 \rho_3 \cr }\right\},
 \nonumber\\
   & &  \nonumber\\
   & &  \nonumber\\
   \left\{ \matrix{ (a_2-b_2)\cr (a_2+b_2)\cr (c_2-d_2)\cr 
                   -(c_2+d_2)}\right\} &=& 
   \int_{-\infty}^{\infty}\frac{d\omega_1}{2\pi}\frac{d\omega_2}{2\pi} 
          a_1^-a_3^+ 
   \left\{\matrix{ \rho_1-\rho_2 \cr \tilde{N}_1(\rho_1-\rho_2) \cr 
                   \rho_3-\rho_4\cr \tilde{N}_1(\rho_3-\rho_4)} \right\},
 \nonumber\\
  & &  \nonumber\\
  & &  \nonumber\\
  \left\{ \matrix{ (a_3-b_3)\cr (a_3+b_3)\cr (c_3-d_3)\cr 
                  -(c_3+d_3)}\right\} &=& 
  \int_{-\infty}^{\infty}\frac{d\omega_1}{2\pi}\frac{d\omega_2}{2\pi} 
        a_1^-a_2^+ 
  \left\{\matrix{ \rho_1-\rho_2 \cr \tilde{N}_1 (\rho_1-\rho_2)\cr 
                  \rho_3-\rho_4\cr \tilde{N}_1 (\rho_3-\rho_4)} \right\}, 
 \nonumber\\
  & &  \nonumber\\
  & &  \nonumber\\
  \left\{ \matrix{ (a_4-b_4)\cr (a_4 + b_4)\cr (c_4-d_4)\cr 
                  -(c_4+d_4)\cr} \right\} &=&  
  \int_{-\infty}^{\infty}\frac{d\omega_1}{2\pi}\frac{d\omega_2}{2\pi} 
        a_2^+a_3^- 
  \left\{ \matrix{ \rho_1 \cr -\tilde{N}_3 \rho_1 \cr \rho_3 \cr 
                  -\tilde{N}_3 \rho_3 \cr }\right\},
 \nonumber\\
  & &  \nonumber\\
  & &  \nonumber\\
  \left\{ \matrix{ (a_5-b_5)\cr (a_5 + b_5)\cr (c_5-d_5)\cr 
                  -(c_5+d_5)\cr} \right\} &=&  
  \int_{-\infty}^{\infty}\frac{d\omega_1}{2\pi}\frac{d\omega_2}{2\pi} 
        a_1^+a_2^- 
  \left\{ \matrix{ \rho_2 \cr -\tilde{N}_2  \rho_2 \cr 
                   \rho_4 \cr  -\tilde{N}_2 \rho_4 \cr }\right\},
 \nonumber\\
  & &  \nonumber\\
  & &  \nonumber\\
  \left\{ \matrix{ (a_6-b_6)\cr (a_6 + b_6)\cr (c_6-d_6)\cr 
                  -(c_6+d_6)\cr} \right\} &=&  
  \int_{-\infty}^{\infty}\frac{d\omega_1}{2\pi}\frac{d\omega_2}{2\pi} 
        a_2^-a_3^+ 
  \left\{ \matrix{ \rho_2 \cr -\tilde{N}_2\rho_2\cr \rho_4 \cr 
                  -\tilde{N}_2 \rho_4 \cr }\right\}. 
 \label{A8}
 \end{eqnarray}

Not all of the spectral functions Eq.~(\ref{eq: spectralfunctions}) are 
independent.  We can define the two independent combinations Eq.~(\ref{AB}) 
and express all results in terms of $\rho_A$ and $\rho_B$ by using 
the relations
 \begin{eqnarray}
   \rho_1-\rho_3 = \tilde{N}_1(\rho_A-\rho_B) - \tilde{N}_2\rho_B,
 \nonumber\\
   \rho_2-\rho_4 = \tilde{N}_1(\rho_A-\rho_B) + \tilde{N}_3\rho_A.
 \label{eq: rhorelations}
 \end{eqnarray}
Using Eqs.~(\ref{A7}),~(\ref{A8}) and~(\ref{eq: rhorelations}) we obtain
Eq. (\ref{eq: spectral}) for the spectral representation of the vertex 
functions.

%
%

\begin{figure}[h]\epsfxsize=7cm\epsfysize=3cm 
\centerline{\epsfbox{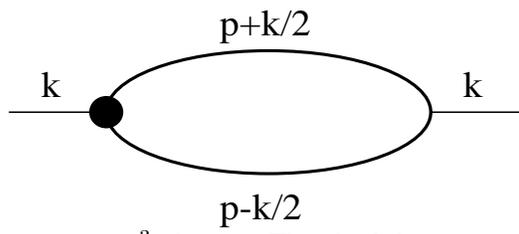}}
\caption{\it The scalar self energy in $\phi^3$ theory. The thick lines 
represent full propagators and the solid blob represents a full 
three-point vertex.
}\label{F1}
\end{figure}

\end{document}